\documentclass[preprint,12pt,prd,amsmath,amssymb,floatfix]{revtex4}

\usepackage{epsfig}
\usepackage{bbold}
\usepackage{bm}
\usepackage{color}
\usepackage{dcolumn}
\usepackage{feynmf} 
\usepackage{graphics}
\usepackage{graphicx}
\usepackage{subfigure}
\usepackage{slashed}
\usepackage{placeins}

\usepackage{pdfpages}
\usepackage{eso-pic}
\usepackage{everyshi}
\usepackage{pdflscape}
\usepackage{amsmath}
\usepackage{dashbox}

\def\be{\begin{equation}}
\def\ee{\end{equation}}
\def\bea{\begin{eqnarray}}
\def\eea{\end{eqnarray}}
\def\bse{\begin{subequations}}
\def\ese{\end{subequations}}
\def\bc{\begin{center}}
\def\ec{\end{center}}

\begin{document}

\title{On ultrahigh-energy neutrino-nucleon deep-inelastic scattering and the
Froissart bound}

\author{A.~V.~Kotikov$^1$\footnote{E-mail: kotikov@theor.jinr.ru}, I.~A.~Kotikov$^2$}
\affiliation{$^1$ Bogoliubov Laboratory of Theoretical Physics, Joint Institute for Nuclear Research, 141980 Dubna, Russia\\
  $^2$ Moscow State University, Faculty of Physics,
  119991, Moscow, Russia}

\date{\today}

\begin{abstract}

  A brief review of the results
  for the total cross section $\sigma^{\nu N}$ of ultrahigh-energy neutrino
  deep inelastic scattering on isoscalar nuclear targets is presented. These results are based on simple approximations for $\sigma^{\nu N}$ and are
  compared with the experimental data of
  the IceCube Collaboration.
  The total cross section $\sigma^{\nu N}$ is proportional to the structure function $F_2^{\nu N}(M_V^2/s,M_V^2)$, where $M_V$ is the
  intermediate boson mass and $s$ is square of the energy of the center of mass.
The coefficient  in the front of $F_2^{\nu N}(M_V^2/s,M_V^2)$ depends on the asymptotic behavior of $F_2^{\nu N}$ at low values of $x$
It contains an additional term $\sim \ln{s}$ if $F_2^{\nu N}$ is scaled by the power $\ln(1/x)$.
Therefore, the asymptotic behavior of $F_2^{\nu N}\propto\ln^2(1/x)$ for small $x$ often assumed in the literature already leads to
violation of the Froissart bound for $\sigma^{\nu N}$ .

\end{abstract}


\maketitle

\section{Introduction }

Today, neutrino cross sections at very high energies can be measured using astrophysical origin neutrinos.
Thus, the IceCube Collaboration published measurements for 10 TeV $\leq E_{\nu} \leq 10^4$ TeV
\cite{IceCube:2017roe,IceCube:2020rnc} based on theoretical predictions for such energies.
They can also be compared with the predictions of the Standard Model for the search for new physics.
Also, the IceCube event analysis provides information on the astrophysical neutrino flux as a function of $E_{\nu}$.
For these reasons, it is important to have modern predictions for the shape of the ultrahigh-energy (UHE) cosmic
neutrino cross sections $\sigma^{\nu N}$.

This requires extrapolation to large values of $E_\nu$, for which there are various approaches (see \cite{Illarionov:2011wc} and references therein).
They are based on the successful description of terrestrial data in terms of perturbative QCD and often
contain the Froissart constraint \cite{Froissart:1961ux} on $\sigma^{\nu N}$.
According to the latter, unitarity and analyticity limit of the growth of the total cross section with energy $s$ as $\ln^2s$.

In this short article, we present a brief overview of the results of \cite{Illarionov:2011wc}, where the general formula for
$\sigma^{\nu N}$ was obtained. It is surprisingly compact and correctly explains the asymptotic behavior at high energies,
which makes it ideally suited to the phenomenology of the UHE neutrino.
The cross section is proportional to the DIS structure function (SF) $F_2^{\nu N}(x,Q^2)$, which has a well-known representation
in terms of parton distribution functions (PDFs) in the framework of the parton model (PM) of QCD, where $x$ and the typical
energy scale $Q$ are properly defined in terms of $E_\nu$ and $M_V$ ($V=W,Z$).
We assume
that the available DIS experimental data allow extrapolation to very high values of $E_\nu$ using the appropriate parameterization
for $F_2^{\nu N}$.
The obtained results for  $\sigma^{\nu N}$ are compared with the experimental data \cite{IceCube:2017roe,IceCube:2020rnc} of
  the IceCube Collaboration.

\section{Approach}

We consider  charged current (C) and neutral current (NC) DIS processes,
\be
\nu(k)+N(P) \to  \ell(k^\prime)+X,~~
\nu(k)+N(P) \to  \nu(k^\prime)+X,
\label{eq:dis}
\ee
respectively, where $N=(p+n)/2$ denotes an isoscalar nucleon target of mass $M$, $X$ collects unobservable parts of the final state,
four-momentum assignments are given in parentheses, and we introduce the familiar kinematic variables
\begin{equation}
s=(k+P)^2,\
Q^2=-q^2,\
x=\dfrac{Q^2}{2q\cdot P},\
y=\dfrac{q\cdot P}{k\cdot P},
\end{equation}
where $q=k-k^\prime$.
In the target rest frame we have $s=M(2E_\nu+M)\approx 2ME_\nu$ and $xy=Q^2/(2ME_\nu)$.
In the kinematic regime of interest to us, inclusive  spin-averaged double-differential cross sections of the processes~(\ref{eq:dis})
are determined in a very good approximation by the expression \cite{McKay:1985nz}:
\begin{equation}
\dfrac{d^2\sigma_i^{\nu N}}{dx\,dy} =
 \dfrac{G_F^2ME_{\nu}}{2\pi}K_i
 \left(\dfrac{M_{V}^2}{Q^2+M_V^2}\right)^2
K(y)F_2^{\nu N},
\label{eq:diff}
\end{equation}
where $i={\rm CC},{\rm NC}$, $G_F$ is the Fermi constant, $K(y)=2-2y+y^2$.
In the so-called {\it wee parton} pattern, suitable for the small $x$ regime \cite{Berger:2007ic},
we have $K_{\rm CC}=1$ and $K_{\rm NC}= 1/ 2-x_w+(10/9)x_w^2$, where $x_w=\sin^2\theta_w$ and $\theta_w$ is the weak mixing angle.
Using $x_w=0.231$ \cite{ParticleDataGroup:2022pth} we get $K_{\rm NC}=0.328$.
The contributions of SFs $F_L^{\nu N}$ and $F_3^{\nu N}$ to the r.h.s. of Eq.~(\ref{eq:diff}) are negligible:
$F_L^{\nu N}$ tends to zero as $Q^2$ increases (see, e.g., \cite{Kotikov:1993yw}) while $F_3^{\nu N}$ is essentially determined
by valence quarks.

A detailed examination of the available $\ell N$ DIS data (see, e.g., \cite{:2009wt}) shows that in the
limit $x \to 0$ $F^{\ell N}_2$ exhibits a singular behavior of the form $F^{\ell N}_2(x,Q^2)\simeq x^{-\delta}\tilde{F}^{\ell N}_2(x,Q^2)$,
where $\delta$ is small positive number, while $\tilde{F}^{\ell N}_2$ diverges less strongly than any power of $x$, {\it i.e.},
$\tilde{F}^{\ell N}_2(x,Q^2)/x^{-\lambda}\to0$ as $x\to0$ for any positive number $\lambda$.
Assuming a symmetric quark sea, which suits the low $x$ regime, we have $F_2^{\nu N}(x,Q^2)=(18/5)F_2^{\ell N}(x ,Q^ 2)$, so that
the behavior of $\tilde F^{\ell N}_2$ at low $x$ carries over to $\tilde F^{\nu N}_2$.

Imposing a lower cut $Q_0^2$ on $Q^2$, the total cross sections of the processes~(\ref{eq:dis}) are estimated as
\begin{equation}
\sigma^{\nu N}_i(E_\nu)=
	\dfrac{1}{2ME_{\nu}}\int_{Q_0^2}^{2ME_\nu}dQ^2
	\int_{\hat{x}}^{1}\dfrac{dx}{x}\,\dfrac{d^2\sigma^{\nu N}_i}{dx\,dy},
\label{eq:tot}
\end{equation}
where $\hat x=Q^2/(2ME_\nu)$.
Substituting Eq.~(\ref{eq:diff}) into (\ref{eq:tot}), we get
\be
\sigma^{\nu N}_i(E_\nu)=
	\dfrac{G_F^2}{4\pi} K_i 
	\int_{Q_0^2}^{2ME_\nu}dQ^2
	\left(\dfrac{M_V^2}{Q^2+M_V^2}\right)^2 \,
	\int_{\hat{x}}^{1}\dfrac{dx}{x}
        K\left(\dfrac{\hat{x}}{x}\right)
	F_2^i(x,Q^2).
\label{eq:tot1}
\ee
Using the low-$x$ asymptotic form $F^{\nu N}_2(x,Q^2)\simeq x^{-\delta}\tilde F^{\nu N}_2(x,Q^2)$
explained above, the inner integral on the r.h.s of Eq.~(\ref{eq:tot1}) can be rewritten as the Mellin convolution
$K(\hat{x})\otimes F^{\nu N}_2(\hat{x}, Q^2)$, which can be represented for small $\hat{x}$ values, in the factorized
form $\tilde M(\hat{x},Q^2,1+\delta)F_2^{\nu N }(\hat{x},Q^2)$ up to terms of ${\cal O}(\hat{x})$ \cite{Lopez:1979bb}.
Here,
\begin{equation}
\tilde M(\hat{x},Q^2,1+\delta)=
2\left(\dfrac{1}{\tilde{\delta}(\hat{x},Q^2)}-\dfrac{1}{\delta}\right)
+M(1+\delta),
\end{equation}
where
\begin{equation}
\dfrac{1}{\tilde{\delta}(x,Q^2)}=
	\dfrac{1}{\tilde F^{\nu N}_2(x,Q^2)}
	\int_x^1 \dfrac{dy}{y}\tilde{F}^{\nu N}_2(y,Q^2)
\label{eq:delta}
\end{equation}
and $M(1+\delta)$ is the analytic continuation of the Mellin moment
\begin{equation}
M(n) = \int_0^1 dx \, x^{n-2} K(x) =
	\dfrac{2}{n-1}-\dfrac{2}{n}+\dfrac{1}{n+1}
\end{equation}
for integer values of $n$.
Hence, Eq.~(\ref{eq:tot1}) becomes
\be
\sigma^{\nu N}_i(E_\nu) \simeq
	\dfrac{G_F^2}{4\pi} K_i 
	\int_{Q_0^2}^{2ME_\nu}dQ^2
	\left(\dfrac{M_V^2}{Q^2+M_V^2}\right)^2 \,
\tilde M(\hat{x},Q^2,1+\delta)F_2^{\nu N}(\hat{x},Q^2).
\label{eq:tot2}
\ee
Since the $Q^2$-dependence of $F_2^{\nu N}(\hat{x},Q^2)$ and, consequently, $\tilde M(\hat{x},Q^2,1+\ delta )$ is only logarithmic,
the factor $[M_V^2/(Q^2+M_V^2)]^2$ essentially fixes the scale $Q^2=M_V^2$ \cite{Fiore:2003kc}, so Eq,~(\ref{eq:tot2}) simplifies to
\begin{equation}
\sigma^{\nu N}_i(E_\nu)\simeq
	\dfrac{G_F^2}{4\pi} K_i M_V^2
\tilde M(\tilde{x},M_V^2,1+\delta)F_2^{\nu N}(\tilde{x},M_V^2),
\label{eq:tot3}
\end{equation}
where $\tilde x=M_V^2/(2ME_\nu)$.

This is our main formula.
Further simplification depends on the $\delta$ value, and we will
distinguish between two cases:\\
(1) If $\delta$ is not too small, so $\hat{x}^{\delta}\ll\text{const}$,
then the lower limit $\hat{x}$ of the inner integral on the r.h.s. of
Eq.~(\ref{eq:tot1}) can be set to zero, so that
\begin{equation}
\tilde M(\hat{x},Q^2,1+\delta)=M(1+\delta)=
\dfrac{4+3\delta+\delta^2}{\delta(\delta+1)(\delta+2)}
\label{eq:largedelta}
\end{equation}
becomes independent of $\hat{x}$ and $Q^2$.\\
(2) On the other hand, if $\delta\ll1$, then we have
\begin{equation}
\tilde M(\hat{x},Q^2,1+\delta)=\tilde M(\hat{x},Q^2,1)
=\dfrac{2}{\tilde{\delta}(\hat{x},Q^2)}-\dfrac{3}{2}.
\label{eq:smalldelta}
\end{equation}
Note that $\tilde{\delta}$ is determined by the asymptotic behavior of $\tilde F^{\nu N}_2$ for small $x$ values.
For example, if $\tilde F^{\nu N}_2(x,Q^2)\propto\ln^p(1/x)$ for $x\to0$, then $1/\tilde{\delta}(x,Q^2)=\ln(1/x)/(p+1)$ \cite{Kotikov:1998qt}.

\begin{table}[t]
\begin{center}
\begin{tabular}{|cccccc|}
\hline
$i$ & $c_i\times10^{3}$ & $\delta_i\times10^{2}$ & $a_{1i}\times10^{2}$
& $a_{2i}\times10^{3}$ & $b_{i}\times10^{2}$ \\
\hline
 0 & $189.4$ & $10.90$ & $-8.471$ & $12.92$ & $2.689$ \\
 1 & $1.811$ & $6.249$ & $4.190$ & $0.2473$ & $11.63$ \\
 2 & $-0.6054$ & $-0.3722$ & $-0.3976$ & $1.642$ & $-0.7307$ \\
\hline
\end{tabular}
\end{center}
\caption{%
The values of the fit parameters appearing in Eqs.~(\ref{CTEQ6:d}), (\ref{BBT:F2.1}), and (\ref{Haidt:F2a}).}
\label{Tab:CTEQ6}
\end{table}

\section{Parameterization of $F_2^{\ell N}$}

In Ref. \cite{Illarionov:2011wc}, Eq.~(\ref{eq:tot3}) was applied to the three most popular types of $F_2^{\ell N}$ parameterizations,
namely the standard PM representation implemented using
proton PDFs \cite{:2009wt,Martin:2009bu,Lai:2010vv}; to a modification of
the simple logarithmic form proposed by Haidt (H) \cite{Haidt:1999ps} and to a more complicated form introduced by Berger, Block and Tan
(BBT) \cite{Block:2006dz}.
While the $Q^2$ dependence of the PM representation of  $F_2^{\nu N}$ is determined by the DGLAP evolution, the $Q^2$ dependences of the H and
BBT forms are directly determined by global fitss of experimental data covering a wide range of $Q^2$ values.

In the range of small $x$ values, the PM parametrization of $F^{\ell N}_2$ can be well approximated by the following ansatz: 
\begin{equation}
F^{\ell N}_{2,\text{PM}}(x,Q^2)=C_\text{PM}(Q^2)x^{-\delta_\text{PM}(Q^2)},
\label{n8.1}
\end{equation}
with
\be
C_\text{PM}(Q^2) = c_0 + c_1 \ln{Q}^2 + c_2 \ln^2{Q}^2,~~
\delta_\text{PM}(Q^2) = \delta_0 + \delta_1 \ln{Q}^2 + \delta_2 \ln^2{Q}^2, 
\label{CTEQ6:d}
\ee
where it is understood that $Q^2$ is taken in units of GeV${}^2$.
Fitting Eqs.~(\ref{n8.1}) and (\ref{CTEQ6:d}) to the result for $F_2^{\ell N}$ evaluated in next-to-leading order (NLO) using
HERAPDF1.0 \cite{:2009wt} set of proton PDFs, the  $c_i$ and $\delta_i$ values were obtained in \cite{Illarionov:2011wc},
where the cut $Q^2>3.5$~GeV$^2$ was imposed to suppress higher-twist effects. The results are collected in Table~\ref{Tab:CTEQ6}.
From Eq.~(\ref{CTEQ6:d}) and Table~\ref{Tab:CTEQ6} we get that
\begin{equation} 
\delta_\text{PM}(M_Z^2)\approx\delta_\text{PM}(M_W^2)\approx0.37,
\label{eq:cteqdelta}
\end{equation}
so Eq.~(\ref{eq:tot3}) must be used with Eq.~(\ref{eq:largedelta}).
Using the MSTW \cite{Martin:2009bu} and CT10 \cite{Lai:2010vv} PDFs, $\delta_\text{PM}(M_V^2)\approx0.35$ and 0.38 were obtained, respectively,
in Ref. \cite{Illarionov:2011wc}.
The resulting high-$E_\nu$ behavior $\sigma^{\nu N}_i(E_\nu)\propto \tilde{x}^{-\delta_\text{PM}(M_V^2)}$ is in good agreement with other studies
\cite{Gandhi:1998ri}.

Here we recall the part of theBBT parameterization of $F_2^{\ell N}$ suitable for the range $x<x_P=0.11$ \cite{Block:2010ud} and needed
for our applications. \footnote{%
We checked that the contribution to Eq.~(\ref{eq:tot1}) from
$F^{\ell N}_{2,\text{BBT}}$, valid in the range $x_P<x<1$,  is numerically
insignificant, according to Refs.~\cite{Block:2006dz,Block:2010ud}.}. 
It reads \cite{Block:2006dz,Block:2010ud}:
\be
F^{\ell N}_{2,\text{BBT}}(x,Q^2) = (1-x)
	\left[A_0 + A_1(Q^2) \ln\dfrac{x_P(1-x)}{x(1-x_P)}
          +A_2(Q^2) \ln^2\dfrac{x_P(1-x)}{x(1-x_P)}\right],
\label{BBT:F2}
\ee
where $A_0=F_P/(1-x_P)$, with $F_P=0.413$ \cite{Block:2010ud}, and  
\begin{equation}
A_i(Q^2)=a_{i0} + a_{i1} \ln{Q^2} + a_{i2} \ln^2{Q^2}\quad(i=1,2),
\label{BBT:F2.1}
\end{equation}
with the $a_{ij}$ values form Table~\ref{Tab:CTEQ6}.
Here Eq.~(\ref{eq:tot3}) is to be used with Eq.~(\ref{eq:smalldelta}) and
we find
\begin{eqnarray}
\dfrac{1}{\tilde{\delta}_\text{BBT}(x,Q^2)}\!&\!\simeq\!&\!
	\dfrac{\sum_{i=0}^2A_i\ln^{i+1}(x_P/x)/(i+1)}
	      {\sum_{i=0}^2A_i\ln^i(x_P/x)}
\simeq \dfrac{1}{3} \ln\dfrac{x_P}{x}.
\nonumber\\
\label{BBT:d}
\end{eqnarray}
From Eqs.~(\ref{BBT:F2}) and (\ref{BBT:d}) we get that in the high-energy limit $s\to\infty$,
$F^{\ell N}_{2,\text{BBT}}(\tilde{x},M_V^2)\propto\ln^2s$ and $1/\tilde{\delta}_\text{BBT}(\tilde{x}, M_V^2)\propto\ln s$.
This brings us to the important observation that $\sigma_{\text{BBT}}^{\nu N}\propto\ln^3s$, which clearly violates the Froissart
bound \cite{Froissart:1961ux} in contrast to the fact that was listed in Refs.~\cite{Block:2006dz,Block:2010ud}.
This violation of the Froissart bound is explained by the presence of the $\sim \ln^2x$ term in Eq.~(\ref{BBT:F2}).

On the other hand, if $F_2^{\ell N}$ increases linearly with $\ln x$ as $x\to0$, then $\sigma_i^{\nu N}\propto\ln^2s$ is in according
to Froissart's constraint.
In fact, this is true for the original H ansatz \cite{Haidt:1999ps}: $B\ln(x_0/x)\ln(1+Q^2/Q_0^2)$, which contains only three free parameters.
To improve the quality of the fits, authors of Ref. \cite{Illarionov:2011wc} introduced three additional parameters:
\be
F^{\ell N}_{2,\text{H}}(x,Q^2) = B_0 + B_1(Q^2) \ln \dfrac{x_0}{x},~~
B_1(Q^2)=\sum_{i=0}^2 b_i \ln^i \left(1+\dfrac{Q^2}{Q^2_0}\right).
\label{Haidt:F2a}
\ee
Eq.~(\ref{eq:tot3}) should be used  again with Eq.~(\ref{eq:smalldelta}) and we get
\begin{equation}
\dfrac{1}{\tilde{\delta}_\text{H}(x,Q^2)} \simeq
\dfrac{\sum_{i=0}^1B_i\ln^{i+1}(x_0/x)/(i+1)}
	      {\sum_{i=0}^1B_i\ln^i(x_0/x)}
\simeq \dfrac{1}{2} \ln\dfrac{x_0}{x},
\label{Haidt:d}
\end{equation}
so $\sigma_\text{H}^{\nu N}\propto\ln^2s$ as it should be.
Fiting Eq.~(\ref{Haidt:F2a}) to a recent combination \cite{:2009wt} of the full H1 and ZEUS datasets on $F_2^{\ell N}$
with the cuts $x<0.01$ and $Q^2>3.5$~GeV$^2$ (see \cite{Illarionov:2011wc}) gives $x_0=0.05791$, $Q^2_0=2.578$~GeV${}^2$, $B_0=0.1697$
and the $b_i$ values given in Table~\ref{Tab:CTEQ6}.

\section{UHE neutrinos}

Now consider $\nu N$ DIS with UHE neutrinos.
Following \cite{Illarionov:2011wc}, we focus on the CC DIS. The corresponding NC results can be obtained by
replacing $K_\text{CC}\to K_\text{NC}$ and $M_W\to M_Z$ in our formulas.
To determine the applicability range of our main formula (\ref{eq:tot3}) for $\sigma_\text{CC}^{\nu N}$ we compare it with the
exact formula (\ref{eq:tot1}) which requires a two-dimensional numerical integration, for the above cases PM, BBT and H.
In each case, we find an excellent match for $E_\nu$ values of $10^7$~GeV and above, which corresponds to $x$ values of $10^{-3}$
and below in $F_2^{\ell N}$.
This is illustrated for the BBT and H cases in Fig.~\ref{Fig2}, where the application of the basic formula (\ref{eq:tot3}), shown by solid lines,
is compared with the application of the exact Eq.~(\ref{eq:tot1}) shown with dotted lines.
The approximation based on Eq.~(\ref{eq:tot3}) can also be slightly improved by calculating $\tilde{\delta}(x,Q^2)$ using one-dimensional
integration according to Eq.~(\ref{eq:delta}) instead of using Eqs.~(\ref{BBT:d}) and (\ref{Haidt:d}). This is shown with dotted lines.

The PM results for $\sigma_\text{CC}^{\nu N}$ are estimated by our main formula (\ref{eq:tot3}) with Eqs.~(\ref{eq:largedelta})
and (\ref{eq:cteqdelta}) are also shown in fig.~\ref{Fig2}.
Comparing them with the corresponding BBT and H results, we see that all three predictions agree relatively well in the
$10^7$~GeV${}\alt E_\nu\alt10^9$~GeV range, where approximations for high $E_ { \nu}$ values are already working, and the
corresponding $F_2^{\ell N}$ parameterizations are still defined by the HERA data.
However, these three predictions steadily diverge as $E_\nu$ further increases until they differ by 1-2 orders of magnitude at
typical values of UHE $E_\nu$, which reflects the different low-$x$ behavior of the corresponding parametrizations of $F_2^{\ell N}$.

 Experimental data \cite{IceCube:2017roe,IceCube:2020rnc}
\footnote{Fig.~\ref{Fig2} shows the experimental data obtained using the so-called Frequentist analysis (see Ref. \cite{IceCube:2020rnc}).
  Data based on so-called Bayesian analysis has slightly larger uncertainties and is not shown. They can be found in
  Ref.~\cite{IceCube:2017roe,IceCube:2020rnc}.}
of the IceCube collaboration for $E_{\nu}\sim 10^{5}\div 10^6$ GeV correspond to $x\sim 10^{-2}\div 10^{-1}$, where
the approximation (\ref{eq:tot3}) is at the limit of its applicability, especially in the BBT and H cases.
 The experimental data \cite{IceCube:2017roe,IceCube:2020rnc} are in good agreement with the results obtained in the PM case. In the BBT and H cases,
 there is good agreement only for the results based on the exact formula (\ref{eq:tot1}) and also for the improved results (\ref{eq:delta})
 for $\tilde{\delta}(x,Q^2)$.

\section{Conclusion}

We have shown the results of \cite{Illarionov:2011wc}, where new compact relations given by Eqs.~(\ref{eq:tot3})--(\ref{eq:smalldelta})
were obtained between the total cross section $\sigma_i^{\nu N}(E_\nu)$ in high $E_\nu$ limit and SF $F_2^{\ell N}(x,Q^2)$ for small $x$ values.
This is especially useful for UHE neutrino physics applications, providing reliable predictions in a very fast and convenient way.
Given in terms of a closed analytic formula ~(\ref{eq:tot3}), it also makes it possible to uniquely determine whether the $\sigma_i^{\nu N}$
obtained for a given functional form $F_2^{\ell N}$, obeys Froissart bound \cite{Froissart:1961ux} or not.
In particular, if for small $x$ values of $F_2^{\ell N}\propto\ln^p(1/x)$, which corresponds to $F_2^{\nu N}(\tilde{x},M_V^2)\propto\ln^ps$ for high $s$
in Eq.~(\ref{eq:tot3}), then the coefficient $\tilde{M}$ in this equation gives an additional factor $\propto\ln s$, so the Froissart bound
is violated for $p>1$.
In fact, this refers to the BBT parameterization \cite{Block:2006dz,Block:2010ud} $F_2^{\ell N}$ for which $p=2$.
On the other hand, H parameterization \cite{Haidt:1999ps} is characterized by $p=1$, so the Froissart bound holds.

\begin{figure}
\begin{center}
\includegraphics[width=17.0cm]{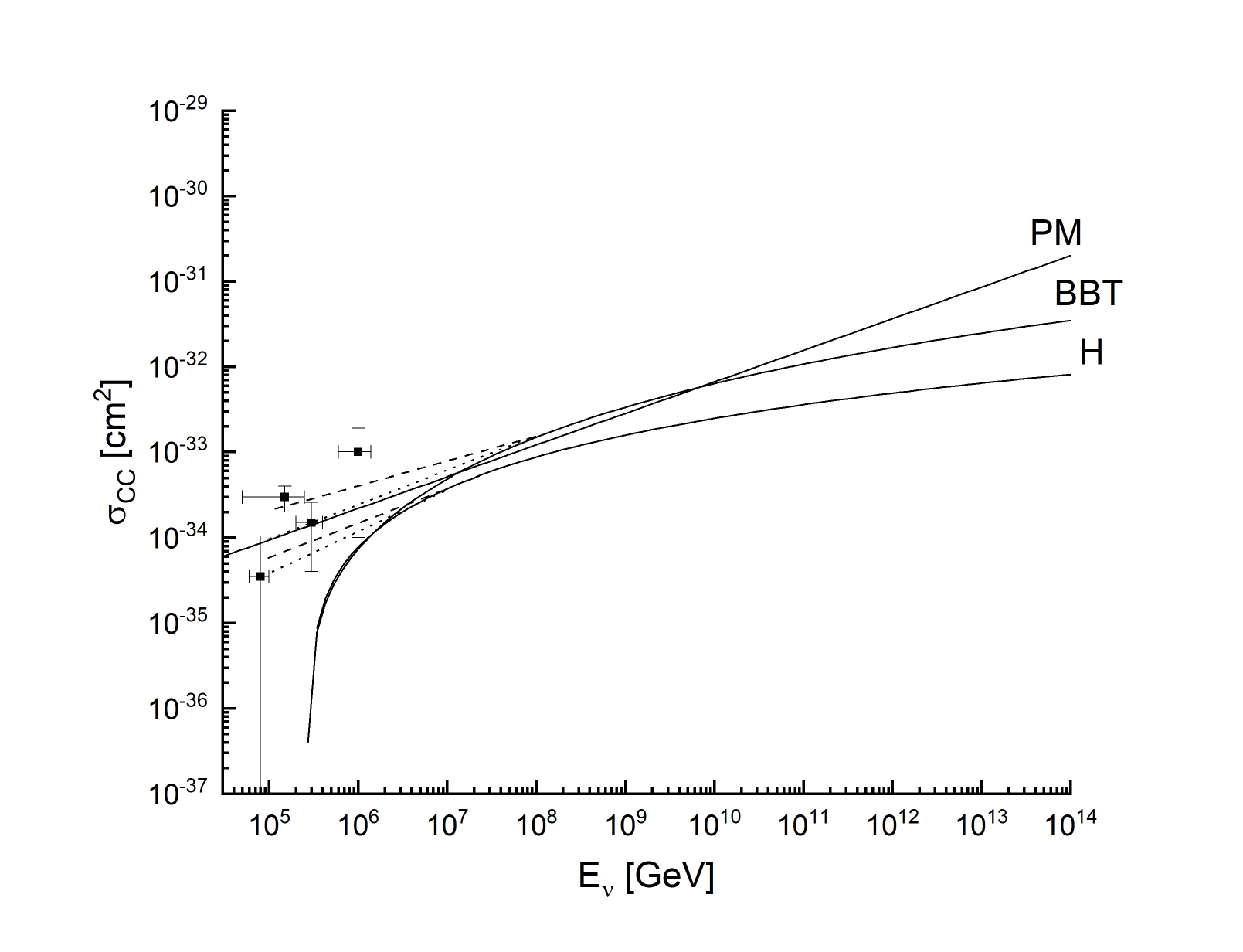}
\end{center}
\caption{
  Predictions for $\sigma_\text{CC}^{\nu N}(E_\nu)$ calculated from the PM, BBT and H  parametrization of $F_2^{\ell N}(x,Q^2)$
  using the main formula (\ref{eq:tot3}) combined with Eqs.~(\ref{eq:largedelta}) or (\ref{eq:smalldelta}) as described in the text.
  In the BBT and H cases, also the improved high $E_\nu$ approximations using Eq.~(\ref{eq:delta}) instead of Eqs.~(\ref{BBT:d})
  and~(\ref{Haidt:d}) (dotted lines) and the exact estimates using Eq.~(\ref{eq:tot1}) (dashed lines) are shown for comparison.
The upper and lower lines correspond to the BBT and H cases, respectively.
The experimental data \cite{IceCube:2017roe,IceCube:2020rnc} of the IceCube collaboration are shown as black dots.
}
\label{Fig2}
\end{figure}

Modern experimental data \cite{IceCube:2017roe,IceCube:2020rnc} of the IceCube collaboration,
obtained at $E_{\nu}\sim 10^{5}\div 10^6$ GeV, are at the limit of applicability of our results (see Fig.~\ref{Fig2}).
The IceCube Collaboration has proposed a major upgrade to the IceCube Antarctic neutrino observatory
(see Ref. \cite{IceCube-Gen2:2020qha} and discussions therein) that will provide measurements of neutrino cross
sections from $E_{\nu} > 10^{11}$ GeV.

Such measurements of $\nu N$ DIS with ultra-high density neutrinos will eventually provide direct access to the
asymptotic behavior of $F_2^{\ell N}$ at small $x$, far beyond the reach of accelerator experiments, and the new relationships
will provide a convenient tool, to open them.
From a theoretical point of view, one important lesson to be learned from our particular example, where total cross sections can simply
be related to structure functions in terms of perturbation theory, is that the direct application of the Froissart constraint to
structure functions presents a potential trap.\\

The work of A.V.K. was supported in part by the Russian
Science Foundation under grant 22-22-00387. He
thanks the Organizing Committee of
the 4th International Symposium on Cosmic Rays and Astrophysics (ISCRA-2023)
 for their invitation.

\end{document}